\documentclass[aps, prb,citeautoscript,reprint,superscriptaddress, nobibnotes, notitlepage]{revtex4-1} 

\usepackage{graphicx}
\usepackage{units}
\usepackage[intlimits]{amsmath}
\usepackage{amssymb}
\usepackage{hyperref}
\usepackage{siunitx}
\usepackage{lineno}
\usepackage{multirow} 
\usepackage{siunitx} 
\usepackage{natbib}
\usepackage[usenames, dvipsnames]{color}
\usepackage[cmyk]{xcolor}
\usepackage{comment}

\usepackage{hyperref}
\usepackage{amsmath}

\newcommand{\beginsupplement}{%
        \setcounter{table}{0}
        \renewcommand{\thetable}{S\arabic{table}}%
        \setcounter{figure}{0}
        \renewcommand{\thefigure}{S\arabic{figure}}%
     }

\begin{document}

\title{Generation and control of frequency dependent squeezing via EPR entanglement}

%

\author{Min Jet Yap}
\email[Corresponding author: ]{minjet.yap@anu.edu.au}
\author{Paul Altin}
\author{Terry G. McRae}
\author{Robert L. Ward}
\author{Bram J. J. Slagmolen}
\author{David E. McClelland}

\affiliation{OzGrav, Department of Quantum Science, Research School of Physics, Australian National University, Acton, Australian Capital Territory 2601, Australia}

\maketitle




	
	\textbf{The Standard Quantum Limit (SQL) in interferometric displacement measurement imposes a restriction on the precision of the measurement due to its quantum back-action noise \cite{Braginsky_SQL}. This limit can be surpassed over a broad frequency band by injecting squeezed vacuum states with a frequency dependent (FD) quadrature angle into the measurement system \cite{Kimble}. Here, we demonstrate the generation of FD squeezed vacuum states by utilizing Einstein-Podolsky-Rosen (EPR) entangled states.  The frequency dependence of the squeezed state can be tuned by conditionally filtering the measurement once the two entangled fields are detected. A stable control scheme for the method is demonstrated which can be implemented in future gravitational-wave detectors to improve their astronomical reach. }
	

	The detection of the binary neutron star inspiral GW170817 by the Advanced Laser Interferometer Gravitational Wave Observatory (LIGO) \cite{LIGO} and Advanced Virgo \cite{VIRGO}, followed by counterpart observations across a broad electromagnetic spectrum, marks the dawn of multi-messenger astronomy \cite{GW170817, GW170817_MM}. The discovery rate and potential of gravitational-wave astronomy is limited by the detector's sensitivity. A key challenge in improving the sensitivity of gravitational-wave detectors is to reduce quantum noise which arises due to the quantum nature of light.
	
	Quantum noise is driven by the vacuum fluctuations which enter into the interferometer via the dark readout port \cite{Caves1981}. This affects the interferometer in two ways: imprecision noise, also called shot noise, which limits how well the phase difference between the interferometer arms can be resolved, and backaction noise, also called quantum radiation pressure noise, which imparts a displacement noise on the suspended mirrors due to amplitude fluctuations. A reduction of the shot noise by increasing the optical power in the interferometer results in an increase in the backaction noise, and the trade-off between these two which results in the minimum total noise is known as the SQL \cite{Kimble, Optomechanics, Braginsky_SQL}.
	
	A well established technique to improve the sensitivity of the interferometer is to replace the coherent vacuum fluctuations into the dark port with a squeezed vacuum state \cite{Kimble}. In accordance with the Heisenberg Uncertainty Principle, the variance of a quadrature in a squeezed state can be reduced at the expense of increased variance in the orthogonal quadrature. This allows the quantum noise to be manipulated, reducing shot noise at the expense of increased backaction noise, or vice-versa. Using squeezed vacuum generated from a nonlinear crystal via the parametric down conversion process \cite{Wade2016, Vahlbruch2016, Oelker2016_SQZ}, shot noise reduction has been successfully demonstrated at the GEO-600 \cite{GEO_SQZ} and LIGO Hanford \cite{LIGO_SQZ} detector, and is now in routine operation in current GW detectors..
	
	However, in order to achieve broadband quantum noise reduction across the entire detection band and to surpass the SQL, the squeezed vacuum state injected must have a frequency-dependent (FD) squeezing quadrature \cite{Kimble}. This rotation can be achieved by passing a frequency-independent squeezed state through a dispersive element with a narrow linewidth \cite{Kimble,Mikhailov2006, Ma2014}. For gravitational-wave interferometers, implementation of this technique requires filter cavities with optical linewidth of the scale of 50\,Hz \cite{Kwee2014, Evans2013}. This is experimentally challenging as the low linewidth requirement dictates a long baseline filter cavity in order to limit performance degradation from optical losses \cite{Oelker2016, Chelkowski, Eleonora2016}. In 2017, Ma et.al. proposed a method to generate FD squeezed states without the need of an external filter cavity \cite{Ma2017}. The proposal relies on the use of Einstein-Podolsky-Rosen (EPR) entangled states, and utilizing the interferometer as both a gravitational-wave detector and a filter cavity. A similar proposal using EPR states has also been studied for GEO600 operating in a detuned configuration\cite{Brown2017}.

	In this work, we verify the proposal by experimentally demonstrating the generation of FD squeezed vacuum with EPR entangled states. Rotation of the squeezing quadrature was achieved with a 0.5\,m long optical resonator, and a squeezing phase control scheme was implemented which is scalable for application in gravitational-wave interferometers.


\begin{figure*}
	\centerline{\includegraphics[width=160 mm]{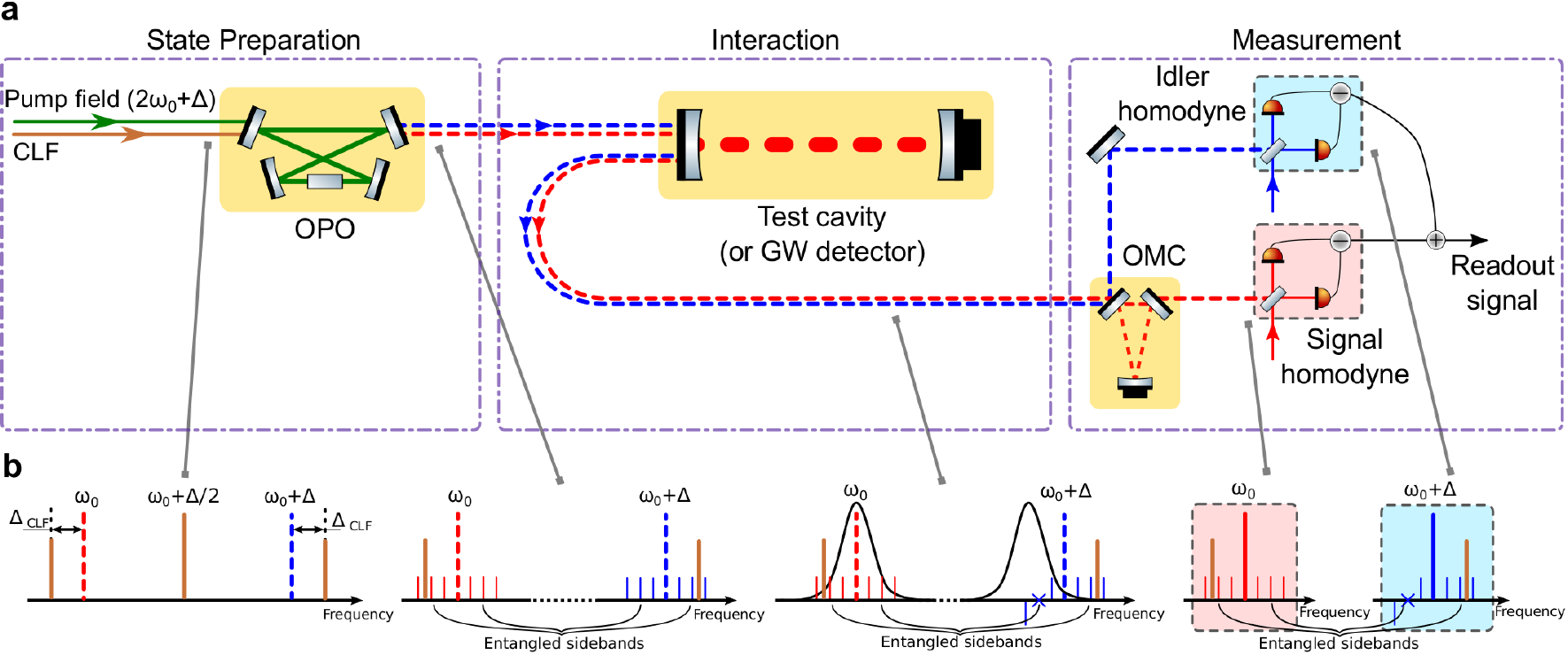}}
	\caption{\textbf{Generation of frequency dependent squeezing with EPR states.} \textbf{a,} Simplified schematic of experiment. The entangled signal (red) and idler (blue) fields generated by the Optical Parametric Oscillator (OPO) are directed towards the test cavity. The reflected fields (which are picked off with a Faraday isolator) are then sent towards the measurement stage where a triangular Output Mode Cleaner (OMC) cavity is used to spatially separate the entangled fields. The two fields are then measured with two homodyne detectors, and are electronically recombined for the final readout measurement. An auxiliary phase modulated Coherent Locking Field (CLF) is injected into the OPO to control the squeezing angle. \textbf{b,} Frequency spacing diagram of the signal, idler and control fields. The OPO pumped at frequency $2\omega_0+\Delta$ entangles sidebands between the signal and idler fields. The sidebands of the idler field experience a differential phase shift when reflected off a detuned cavity. The $\times$ indicates a sideband pointing into the page.}
	\label{fig:Setup}
\end{figure*}


	The scheme works by first generating an EPR entangled signal and idler pair at frequency $\omega_0$ and $\omega_0+\Delta$ respectively, as shown in Figure \ref{fig:Setup}, by pumping an Optical Parametric Oscillator (OPO) at frequency $2\omega_0+\Delta$ \cite{Hage2010, Zhang2003, Li2017}. This results in the signal field upper sideband being correlated with the idler field lower sideband, and vice-versa. The entangled pair are then injected into a test cavity such that the signal field is on-resonance with the cavity, while the idler field is detuned from the cavity. Due to the detuned cavity response, the upper and lower sidebands of the idler fields undergo a frequency dependent differential phase rotation, resulting in a quadrature rotation. In the original proposal \cite{Ma2017}, the entangled fields are injected into the dark port of the gravitational-wave interferometer instead. The reflected fields of the test cavity are then picked off and the signal and idler fields are spatially separated with an additional optical cavity, which acts as a frequency selective beamsplitter, before being measured by two separate homodyne detectors. The measurement of the idler field is combined with the signal field measurement resulting in conditional squeezing in a frequency-dependent manner. In a gravitational-wave detector, the idler field needs to undergo Wiener filtering before the recombination for the FD squeezing to counteract the quantum radiation pressure noise, achieving the broadband reduction of quantum noise.
	
	In our experiment, the OPO produced entangled signal and idler fields with a nominal separation of $\Delta = 851$\,MHz which corresponds to the OPO free spectral range. The test cavity was an over-coupled linear cavity with a length of L $\approx$ $0.5$\,m and a half-width-half-maximum (HWHM) of $\gamma_{tc}$ = 1.25\,MHz. In the final measurement, the two homodyne readouts are combined passively with an electronic combiner where the measurement spectrum is taken, and the power of the signal and idler LO is tuned to ensure that the readouts are combined equally. A more detailed description of the experiment is found in the Supplementary Information.

\begin{figure*}
	\centerline{\includegraphics[width=160 mm]{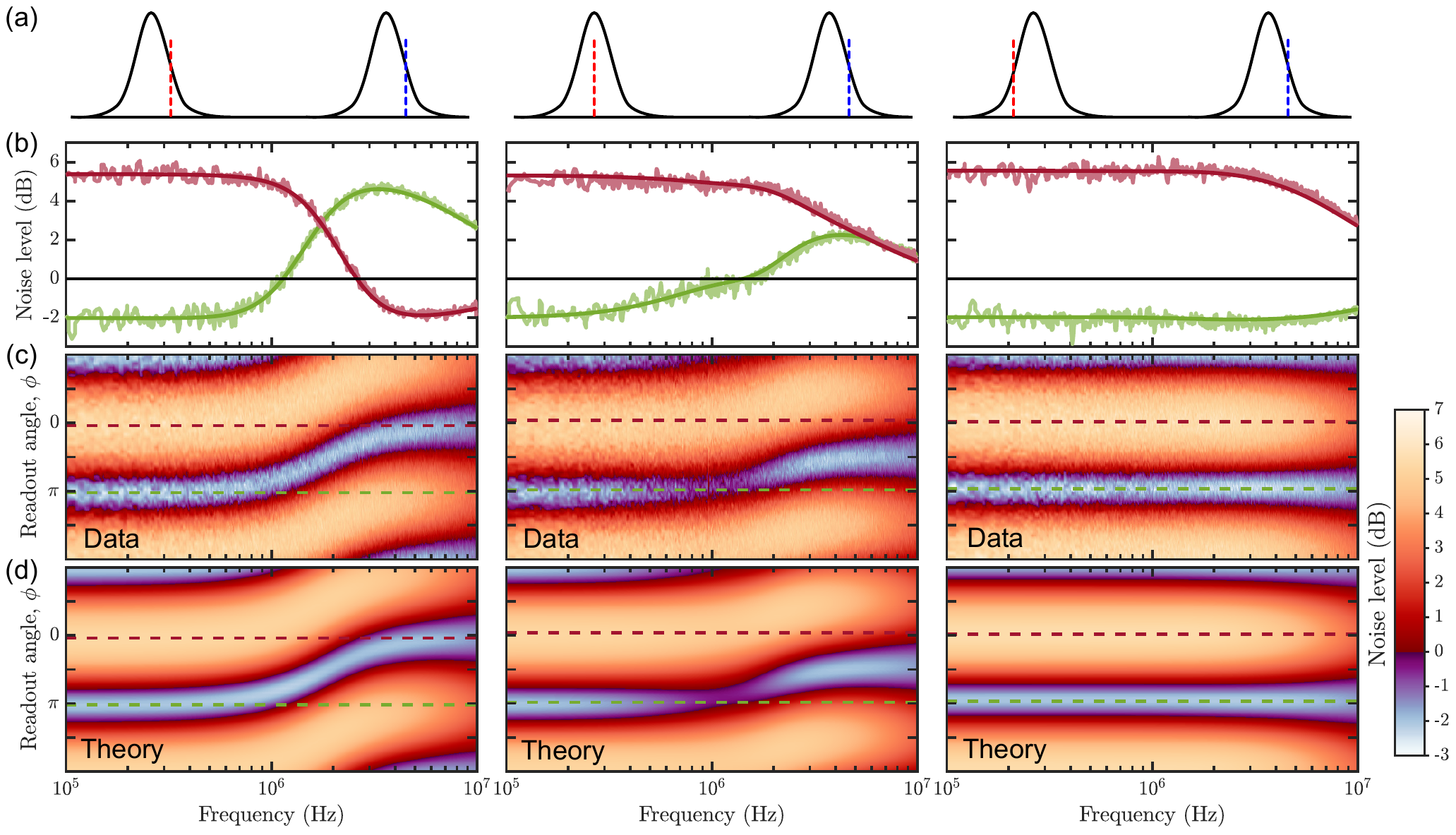}}
	\caption{\textbf{Noise spectrum for different test cavity detuning.} \textbf{a,} Frequency spacing diagram of the signal (red) and idler (blue) fields with respect to test cavity resonances for $\delta_{sig} = \delta_{idl} = \gamma_{tc}$ (left), $\delta_{sig} = 0$, $\delta_{idl} = \gamma_{tc}$ (middle), and $\delta_{sig} = -\delta_{idl} = 0.5\gamma_{tc}$ (right). \textbf{b,} Noise spectrum at the combined output measured at readout angle 0 and $\pi$ (Maroon and green respectively). Solid dark colour lines are fits to the theoretical model. \textbf{c,} Measured noise spectrum as a function of readout angle. \textbf{d,} Associated modelled noise spectrum from the theoretical model. Dotted lines in \textbf{c} and \textbf{d} correspond to the spectra in \textbf{b}. } 
	\label{fig:FDSQZ}
\end{figure*}
	
	
	Reflection from the detuned test cavity imparts a quadrature rotation on the entangled fields. This results in frequency dependent squeezing being generated at the combined spectrum. Experimentally measured noise spectra of the combined output are shown in Figure \ref{fig:FDSQZ} for different signal and idler detuning, $\delta_{sig/idl}$ from the test cavity resonance. The measured data is acquired by recording a spectrogram of the combined output while a linear ramp is placed on the idler LO phase. The electronic dark noise which is approximately 10\,dB below the combined shot noise level was also subtracted from the final measurement.

	The case for $\delta_{sig} = \delta_{idl} = \gamma_{tc}$ is shown in the left column of Figure \ref{fig:FDSQZ}. The same quadrature rotation is imparted onto both the entangled fields which results in the output spectrum being rotated to the orthogonal quadrature about $\gamma_{tc}$. This is equivalent to a squeezed state reflecting off the test cavity with detuning $\gamma_{tc}$. The observed 2\,dB reduction below shot noise correspond to a total loss of $47\pm 4\%$ based on fits from the theoretical model. The high loss is predominately due to the low quantum efficiency of the homodyne photodiodes ($80\%$) used in the experiment. A separate loss measurement was performed to reconfirm this and is described in the Methods section. The reduction in the squeezing and anti-squeezing level at high frequency is due to the OPO HWHM of $\gamma_{OPO} = 12.1$\,MHz.
	
	The middle column of Figure \ref{fig:FDSQZ} shows the noise spectrum of the combined output when $\delta_{sig} = 0$ and $\delta_{idl} = \gamma_{tc}$ as proposed in the original paper \cite{Ma2017}. The quadrature of the idler field is rotated in a frequency-dependent manner by the detuned cavity, and is conditioned onto the signal field resulting in the FD squeezing being produced. The optimal rotation to the orthogonal quadrature can be achieved by implementing Wiener filtering when combining the spectrum \cite{Ma2017}.
	
	Operating a gravitational-wave interferometer with a detuned signal recycling cavity allows for enhanced strain sensitivity at higher measurement frequencies \cite{Buonanno2001}. The detuned cavity however rotates the optimal readout quadrature, requiring FD squeezing to achieve a broadband reduction in quantum noise around the detuned frequency \cite{Brown2017}. By detuning the entangled fields at opposite sides of the test cavity, the quadrature rotation incurred by the entangled fields cancels one another. This results in no frequency dependent squeezing being observed, as shown in the right column of Figure \ref{fig:FDSQZ} where $\delta_{sig} = -\delta_{idl} = 0.5\gamma_{tc}$.


\begin{figure}
	\centerline{\includegraphics[width=85 mm]{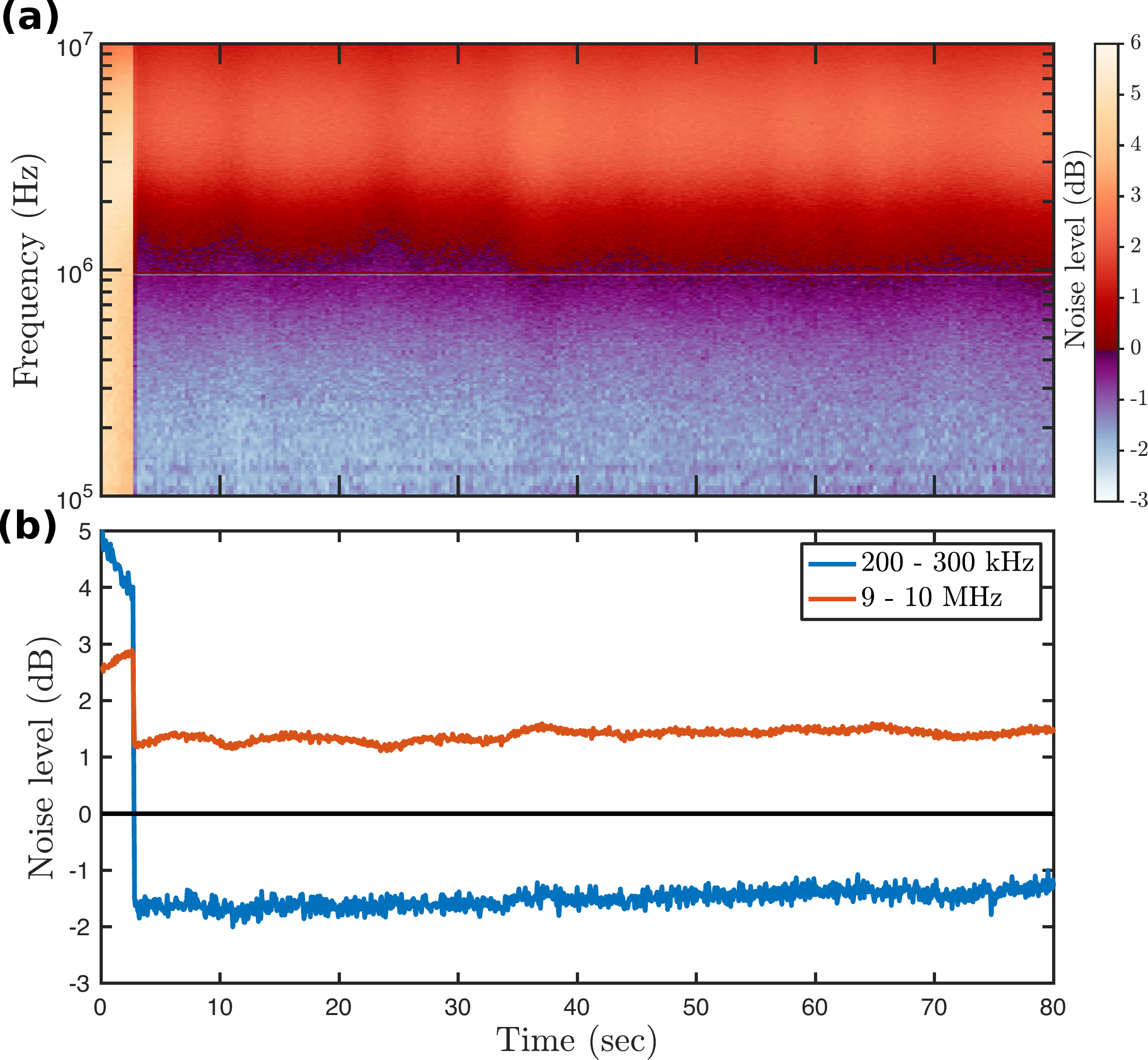}}
	\caption{\textbf{Stable control of the squeezing angle.} \textbf{a,} Spectrogram of the combined output for $\delta_{sig} = 0$, $\delta_{idl} = \gamma_{tc}$. Squeezing angle is stabilized when the CLF lock is engaged at t = 3 sec.  \textbf{b,} Noise level from the spectrogram averaged between 200 - 300\,kHz (blue), and 9 - 10\,MHz (orange).} 
	\label{fig:CLF}
\end{figure}

	To achieve stable phase control of the squeezed ellipse, an auxiliary Coherent Locking Field (CLF) was introduced \cite{Henning, Chua11}. A phase modulated CLF at half the OPO pump frequency is injected into the OPO to generate three error signals for controlling the readout angle, as shown in Figure \ref{fig:Setup}. The parametric down conversion process in the OPO imparts a phase shift on the CLF sidebands, allowing the CLF to track the OPO pump field phase via active feedback to the CLF phase. The transmitted CLF lower (upper) sideband then co-propagates with the entangled fields and interferes with the signal (idler) homodyne LO with a beat note frequency $\Delta_{CLF} = 12.07$\,MHz. The error signal generated from this beat note is fed back to both the signal and idler LO phase, allowing the readout angle to be stabilized. Derivation of the CLF error signals are presented in the Supplementary Information.

	Figure \ref{fig:CLF} shows a spectrogram of the frequency dependent squeezing measurement with the CLF stabilization engaged with $\delta_{sig} = 0$ and $\delta_{idl} = 1.01\gamma_{tc}$. Stable lock was maintained for approximately 80 seconds before the readout angle started to drift. This drift was attributed to drifts in the error signal offset of the CLF lock at the OPO reflection due to residual amplitude modulation from the Electro-Optic Modulator (EOM) used to generate the CLF sideband. The CLF stabilization can be improved by slightly increasing the transmission of the mirror through which the CLF is injected into the OPO in order to improve the signal-to-noise ratio of the CLF error signal. Implementation of active length stabilization for the mode cleaner cavity would also improve the long term stability of the CLF stabilization system.


	The results presented here demonstrate the feasibility of generating FD squeezed vacuum with EPR entangled states, and a scalable approach for implementation into gravitational-wave detectors. For a given amount of squeezed vacuum injected into the interferometer, the EPR approach experiences a signal-to-noise ratio reduction of a factor of two compared to traditional schemes using an auxiliary filter cavity due to additional noise contribution from the idler field \cite{Ma2017}. However, this technique also obviates the need for a separate suspended, low-loss filter cavity, which is expensive and would require a considerably more complex control scheme. It has long been realised that squeezing is a manifestation of entanglement between sidebands of a degenerate OPO \cite{Hage2010}.  What this work shows is that, with careful design, more general entangled states can be manipulated to provide tunable quantum noise suppression in ways that go beyond standard squeezing.

\section*{Methods}

%

\subsection*{Detection loss estimation}

The amount of squeezed light reduction observed is limited by losses which introduce vacuum fluctuations back into the entangled field, degrading the coherence between the signal and idler fields. The noise variance in the anti-squeezed and squeezed quadrature, $V_{\pm}$ scaled to the combined shot noise limit of the signal and idler field is given by \cite{Takeno07}
\begin{align}
	V_{\pm} = 1 \pm \frac{4 x (1-l)}{(1 \mp x)^2 + (\Omega/\gamma_{OPO})^2}
	\label{eq:loss}
\end{align}
where $l$ is the total detection loss, $x$ is the normalized pump parameter which scales with the OPO pump power, $\gamma_{OPO}$ is the OPO HWHM, and $\Omega$ is the measurement frequency. Based on measurements of the squeezed and anti-squeezed level for different OPO pump power, the total detection loss is estimated to be $49 \pm 2$\,\%. The relative phase fluctuation between the squeezed and readout quadrature would also limit the amount of squeezing observed. However due to the relatively high losses in the system, phase noise does not have a significant impact on the experiment.

\section*{Acknowledgements}
This research was supported by the Australian Research Council under the ARC Centre of Excellence for Gravitational Wave Discovery, grant number CE170100004. M.J.Y. acknowledges Ben Buchler, Georgia Mansell and Vaishali Adya for helpful discussions.

\section*{Author contributions}

M.J.Y.performed the investigation and formal analysis. P.A. wrote the real-time data acquisition/analysis program. M.J.Y. and D.E.M. conceptualized the project. M.J.Y. performed writing - original draft; P.A., T.G.M., R.L.W. and D.E.M performed writing - review and editing. P.A., T.G.M., R.L.W. B.J.J.S, and D.E.M provided supervision.

\section*{Competing Interests}
The authors declare no competing interests.

---
%
%
%



\begin{thebibliography}{10}
\expandafter\ifx\csname url\endcsname\relax
  \def\url#1{\texttt{#1}}\fi
\expandafter\ifx\csname urlprefix\endcsname\relax\def\urlprefix{URL }\fi
\providecommand{\bibinfo}[2]{#2}
\providecommand{\eprint}[2][]{\url{#2}}

\bibitem{Braginsky_SQL}
\bibinfo{author}{Braginsky, V.~B.}
\newblock \bibinfo{title}{Classical and quantum restrictions on the detection
  of weak disturbances of a macroscopic oscillator}.
\newblock \emph{\bibinfo{journal}{Sov. J. Exp. Theor. Phys.}}
  \textbf{\bibinfo{volume}{26}}, \bibinfo{pages}{831} (\bibinfo{year}{1968}).

\bibitem{Kimble}
\bibinfo{author}{Kimble, H.~J.}, \bibinfo{author}{Levin, Y.},
  \bibinfo{author}{Matsko, A.~B.}, \bibinfo{author}{Thorne, K.~S.} \&
  \bibinfo{author}{Vyatchanin, S.~P.}
\newblock \bibinfo{title}{Conversion of conventional gravitational-wave
  interferometers into quantum nondemolition interferometers by modifying their
  input and/or output optics}.
\newblock \emph{\bibinfo{journal}{Phys. Rev. D}} \textbf{\bibinfo{volume}{65}},
  \bibinfo{pages}{022002} (\bibinfo{year}{2001}).

\bibitem{LIGO}
\bibinfo{author}{{The LIGO Scientific Collaboration.}}
\newblock \bibinfo{title}{{Advanced LIGO}}.
\newblock \emph{\bibinfo{journal}{Classical and Quantum Gravity}}
  \textbf{\bibinfo{volume}{32}}, \bibinfo{pages}{074001}
  (\bibinfo{year}{2015}).

\bibitem{VIRGO}
\bibinfo{author}{{The VIRGO Collaboration.}}
\newblock \bibinfo{title}{{Advanced Virgo: a second-generation interferometric
  gravitational wave detector}}.
\newblock \emph{\bibinfo{journal}{Classical and Quantum Gravity}}
  \textbf{\bibinfo{volume}{32}}, \bibinfo{pages}{024001}
  (\bibinfo{year}{2015}).

\bibitem{GW170817}
\bibinfo{author}{The LIGO Scientific Collaboration.} \&
  \bibinfo{author}{VIRGO Collaboration}
\newblock \bibinfo{title}{GW170817: Observation of gravitational waves from a
  binary neutron star inspiral}.
\newblock \emph{\bibinfo{journal}{Phys. Rev. Lett.}}
  \textbf{\bibinfo{volume}{119}}, \bibinfo{pages}{161101}
  (\bibinfo{year}{2017}).

\bibitem{GW170817_MM}
\bibinfo{author}{GROND, SALT Group, OzGrav,CAASTROs, DFN, DES, INTEGRAL, Virgo, Insight-HXMT, MAXI Team, J-GEM, RATIR, ATLAS, IceCube,LWA, ePESSTO, GRAWITA, RIMAS, SKA South Africa/MeerKAT, H.E.S.S., Fermi Large Area Telescope, 1M2HTeam, IKI-GW Follow-up, Fermi GBM, Pi of Sky, DWF (Deeper Wider Faster Program), MASTER, AstroSatCadmium Zinc Telluride Imager Team, Swift, PierreAuger, ASKAP, VINROUGE, JAGWAR, Chandra Team at McGill University, TTU-NRAO, GROWTH, AGILETeam, MWA, ATCA, AST3, TOROS, Pan-STARRS, NuSTAR, BOOTES, CaltechNRAO, LIGO Scientific, High Time Resolution Universe Survey, Nordic Optical Telescope, Las Cumbres Observatory Group, TZAC Con-sortium, LOFAR, IPN, DLT40, Texas Tech University, HAWC, ANTARES, KU, Dark Energy Camera GW-EM,CALET, Euro VLBI Team, ALMA.}
\newblock \bibinfo{title}{Multi-messenger observations of a binary neutron star
  merger}.
\newblock \emph{\bibinfo{journal}{The Astrophysical Journal}}
  \textbf{\bibinfo{volume}{848}}, \bibinfo{pages}{L12} (\bibinfo{year}{2017}).

\bibitem{Caves1981}
\bibinfo{author}{Caves, C.~M.}
\newblock \bibinfo{title}{Quantum-mechanical noise in an interferometer}.
\newblock \emph{\bibinfo{journal}{Phys. Rev. D}} \textbf{\bibinfo{volume}{23}},
  \bibinfo{pages}{1693--1708} (\bibinfo{year}{1981}).

\bibitem{Optomechanics}
\bibinfo{author}{Aspelmeyer, M.}, \bibinfo{author}{Kippenberg, T.~J.} \&
  \bibinfo{author}{Marquardt, F.}
\newblock \bibinfo{title}{Cavity optomechanics}.
\newblock \emph{\bibinfo{journal}{Rev. Mod. Phys.}}
  \textbf{\bibinfo{volume}{86}}, \bibinfo{pages}{1391--1452}
  (\bibinfo{year}{2014}).

\bibitem{Wade2016}
\bibinfo{author}{Wade, A.~R.} \emph{et~al.}
\newblock \bibinfo{title}{Optomechanical design and construction of a
  vacuum-compatible optical parametric oscillator for generation of squeezed
  light}.
\newblock \emph{\bibinfo{journal}{Review of Scientific Instruments}}
  \textbf{\bibinfo{volume}{87}}, \bibinfo{pages}{063104}
  (\bibinfo{year}{2016}).

\bibitem{Vahlbruch2016}
\bibinfo{author}{Vahlbruch, H.}, \bibinfo{author}{Mehmet, M.},
  \bibinfo{author}{Danzmann, K.} \& \bibinfo{author}{Schnabel, R.}
\newblock \bibinfo{title}{Detection of 15 {dB} squeezed states of light and their
  application for the absolute calibration of photoelectric quantum
  efficiency}.
\newblock \emph{\bibinfo{journal}{Phys. Rev. Lett.}}
  \textbf{\bibinfo{volume}{117}}, \bibinfo{pages}{110801}
  (\bibinfo{year}{2016}).

\bibitem{Oelker2016_SQZ}
\bibinfo{author}{Oelker, E.} \emph{et~al.}
\newblock \bibinfo{title}{Ultra-low phase noise squeezed vacuum source for
  gravitational wave detectors}.
\newblock \emph{\bibinfo{journal}{Optica}} \textbf{\bibinfo{volume}{3}},
  \bibinfo{pages}{682--685} (\bibinfo{year}{2016}).

\bibitem{GEO_SQZ}
\bibinfo{author}{Grote, H.} \emph{et~al.}
\newblock \bibinfo{title}{First long-term application of squeezed states of
  light in a gravitational-wave observatory}.
\newblock \emph{\bibinfo{journal}{Phys. Rev. Lett.}}
  \textbf{\bibinfo{volume}{110}}, \bibinfo{pages}{181101}
  (\bibinfo{year}{2013}).

\bibitem{LIGO_SQZ}
\bibinfo{author}{{The LIGO Scientific Collaboration}}.
\newblock \bibinfo{title}{Enhanced sensitivity of the ligo gravitational wave
  detector by using squeezed states of light}.
\newblock \emph{\bibinfo{journal}{Nature Photonics}}
  \textbf{\bibinfo{volume}{7}}, \bibinfo{pages}{613--619}
  (\bibinfo{year}{2013}).

\bibitem{Mikhailov2006}
\bibinfo{author}{Mikhailov, E.~E.}, \bibinfo{author}{Goda, K.},
  \bibinfo{author}{Corbitt, T.} \& \bibinfo{author}{Mavalvala, N.}
\newblock \bibinfo{title}{Frequency-dependent squeeze-amplitude attenuation and
  squeeze-angle rotation by electromagnetically induced transparency for
  gravitational-wave interferometers}.
\newblock \emph{\bibinfo{journal}{Phys. Rev. A}} \textbf{\bibinfo{volume}{73}},
  \bibinfo{pages}{053810} (\bibinfo{year}{2006}).

\bibitem{Ma2014}
\bibinfo{author}{Ma, Y.} \emph{et~al.}
\newblock \bibinfo{title}{Narrowing the filter-cavity bandwidth in
  gravitational-wave detectors via optomechanical interaction}.
\newblock \emph{\bibinfo{journal}{Phys. Rev. Lett.}}
  \textbf{\bibinfo{volume}{113}}, \bibinfo{pages}{151102}
  (\bibinfo{year}{2014}).
\newblock

\bibitem{Kwee2014}
\bibinfo{author}{Kwee, P.}, \bibinfo{author}{Miller, J.},
  \bibinfo{author}{Isogai, T.}, \bibinfo{author}{Barsotti, L.} \&
  \bibinfo{author}{Evans, M.}
\newblock \bibinfo{title}{Decoherence and degradation of squeezed states in
  quantum filter cavities}.
\newblock \emph{\bibinfo{journal}{Phys. Rev. D}} \textbf{\bibinfo{volume}{90}},
  \bibinfo{pages}{062006} (\bibinfo{year}{2014}).

\bibitem{Evans2013}
\bibinfo{author}{Evans, M.}, \bibinfo{author}{Barsotti, L.},
  \bibinfo{author}{Kwee, P.}, \bibinfo{author}{Harms, J.} \&
  \bibinfo{author}{Miao, H.}
\newblock \bibinfo{title}{Realistic filter cavities for advanced gravitational
  wave detectors}.
\newblock \emph{\bibinfo{journal}{Phys. Rev. D}} \textbf{\bibinfo{volume}{88}},
  \bibinfo{pages}{022002} (\bibinfo{year}{2013}).

\bibitem{Oelker2016}
\bibinfo{author}{Oelker, E.} \emph{et~al.}
\newblock \bibinfo{title}{Audio-band frequency-dependent squeezing for
  gravitational-wave detectors}.
\newblock \emph{\bibinfo{journal}{Phys. Rev. Lett.}}
  \textbf{\bibinfo{volume}{116}}, \bibinfo{pages}{041102}
  (\bibinfo{year}{2016}).

\bibitem{Chelkowski}
\bibinfo{author}{Chelkowski, S.} \emph{et~al.}
\newblock \bibinfo{title}{Experimental characterization of frequency-dependent
  squeezed light}.
\newblock \emph{\bibinfo{journal}{Phys. Rev. A}} \textbf{\bibinfo{volume}{71}},
  \bibinfo{pages}{013806} (\bibinfo{year}{2005}).

\bibitem{Eleonora2016}
\bibinfo{author}{Capocasa, E.} \emph{et~al.}
\newblock \bibinfo{title}{Estimation of losses in a 300 m filter cavity and
  quantum noise reduction in the kagra gravitational-wave detector}.
\newblock \emph{\bibinfo{journal}{Phys. Rev. D}} \textbf{\bibinfo{volume}{93}},
  \bibinfo{pages}{082004} (\bibinfo{year}{2016}).

\bibitem{Ma2017}
\bibinfo{author}{Ma, Y.} \emph{et~al.}
\newblock \bibinfo{title}{Proposal for gravitational-wave detection beyond the
  standard quantum limit through {EPR} entanglement}.
\newblock \emph{\bibinfo{journal}{Nature Physics}}
  \textbf{\bibinfo{volume}{13}}, \bibinfo{pages}{776--780}
  (\bibinfo{year}{2017}).

\bibitem{Brown2017}
\bibinfo{author}{Brown, D.~D.} \emph{et~al.}
\newblock \bibinfo{title}{Broadband sensitivity enhancement of detuned
  dual-recycled michelson interferometers with epr entanglement}.
\newblock \emph{\bibinfo{journal}{Phys. Rev. D}} \textbf{\bibinfo{volume}{96}},
  \bibinfo{pages}{062003} (\bibinfo{year}{2017}).

\bibitem{Hage2010}
\bibinfo{author}{Hage, B.}, \bibinfo{author}{Samblowski, A.} \&
  \bibinfo{author}{Schnabel, R.}
\newblock \bibinfo{title}{Towards einstein-podolsky-rosen quantum channel
  multiplexing}.
\newblock \emph{\bibinfo{journal}{Phys. Rev. A}} \textbf{\bibinfo{volume}{81}},
  \bibinfo{pages}{062301} (\bibinfo{year}{2010}).

\bibitem{Zhang2003}
\bibinfo{author}{Zhang, J.}
\newblock \bibinfo{title}{Einstein-podolsky-rosen sideband entanglement in
  broadband squeezed light}.
\newblock \emph{\bibinfo{journal}{Phys. Rev. A}} \textbf{\bibinfo{volume}{67}},
  \bibinfo{pages}{054302} (\bibinfo{year}{2003}).

\bibitem{Li2017}
\bibinfo{author}{Li, W.}, \bibinfo{author}{Jin, Y.}, \bibinfo{author}{Yu, X.}
  \& \bibinfo{author}{Zhang, J.}
\newblock \bibinfo{title}{Enhanced detection of a low-frequency signal by using
  broad squeezed light and a bichromatic local oscillator}.
\newblock \emph{\bibinfo{journal}{Phys. Rev. A}} \textbf{\bibinfo{volume}{96}},
  \bibinfo{pages}{023808} (\bibinfo{year}{2017}).

\bibitem{Buonanno2001}
\bibinfo{author}{Buonanno, A.} \& \bibinfo{author}{Chen, Y.}
\newblock \bibinfo{title}{Quantum noise in second generation, signal-recycled
  laser interferometric gravitational-wave detectors}.
\newblock \emph{\bibinfo{journal}{Phys. Rev. D}} \textbf{\bibinfo{volume}{64}},
  \bibinfo{pages}{042006} (\bibinfo{year}{2001}).

\bibitem{Henning}
\bibinfo{author}{Vahlbruch, H.} \emph{et~al.}
\newblock \bibinfo{title}{Coherent control of vacuum squeezing in the
  gravitational-wave detection band}.
\newblock \emph{\bibinfo{journal}{Phys. Rev. Lett.}}
  \textbf{\bibinfo{volume}{97}}, \bibinfo{pages}{011101}
  (\bibinfo{year}{2006}).

\bibitem{Chua11}
\bibinfo{author}{Chua, S. S.~Y.} \emph{et~al.}
\newblock \bibinfo{title}{Backscatter tolerant squeezed light source for
  advanced gravitational-wave detectors}.
\newblock \emph{\bibinfo{journal}{Opt. Lett.}} \textbf{\bibinfo{volume}{36}},
  \bibinfo{pages}{4680--4682} (\bibinfo{year}{2011}).

\bibitem{Takeno07}
\bibinfo{author}{Takeno, Y.}, \bibinfo{author}{Yukawa, M.},
  \bibinfo{author}{Yonezawa, H.} \& \bibinfo{author}{Furusawa, A.}
\newblock \bibinfo{title}{Observation of -9 {dB} quadrature squeezing with
  improvement of phase stability in homodyne measurement}.
\newblock \emph{\bibinfo{journal}{Opt. Express}} \textbf{\bibinfo{volume}{15}},
  \bibinfo{pages}{4321--4327} (\bibinfo{year}{2007}).

\end{thebibliography}

\begin{thebibliography}{1}

\bibitem{Hage2010}
B.~Hage, A.~Samblowski, and R.~Schnabel, ``Towards einstein-podolsky-rosen
  quantum channel multiplexing,'' {\em Phys. Rev. A}, vol.~81, p.~062301, Jun
  2010.

\bibitem{Drever1983}
R.~W.~P. Drever, J.~L. Hall, F.~V. Kowalski, J.~Hough, G.~M. Ford, A.~J.
  Munley, and H.~Ward, ``Laser phase and frequency stabilization using an
  optical resonator,'' {\em Applied Physics B}, vol.~31, no.~2, pp.~97--105,
  1983.

\bibitem{Collett1984}
M.~J. Collett and C.~W. Gardiner, ``Squeezing of intracavity and traveling-wave
  light fields produced in parametric amplification,'' {\em Phys. Rev. A},
  vol.~30, pp.~1386--1391, Sep 1984.

\bibitem{Gardiner1985}
C.~W. Gardiner and M.~J. Collett, ``Input and output in damped quantum systems:
  Quantum stochastic differential equations and the master equation,'' {\em
  Phys. Rev. A}, vol.~31, pp.~3761--3774, Jun 1985.

\end{thebibliography}

\clearpage

\onecolumngrid

\section*{Supplementary Material}
\beginsupplement

\section{Experimental Setup}

\begin{figure*}[b]
	\centerline{\includegraphics[width=140 mm]{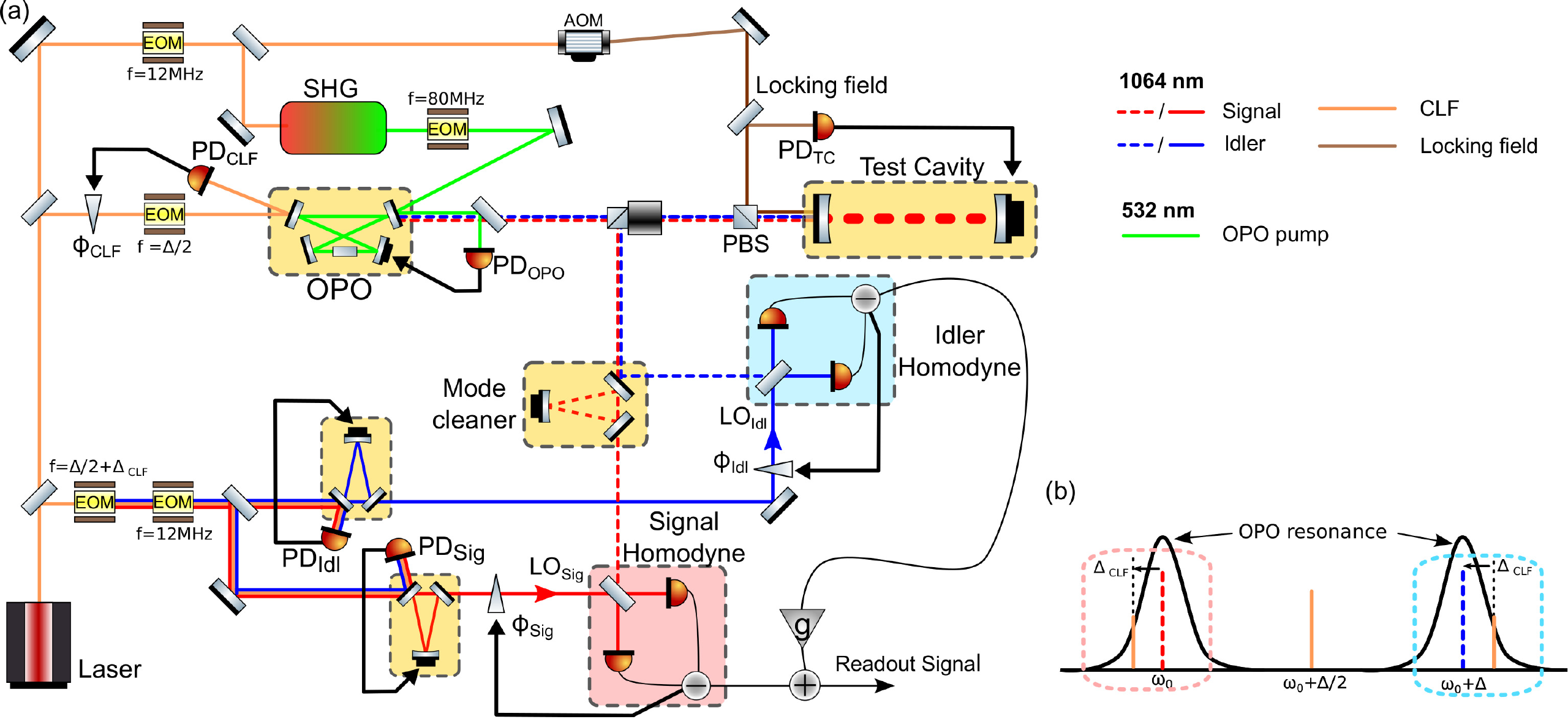}}
	\caption{\textbf{Experimental setup} \textbf{a,} The entangled signal and idler fields generated by the Optical Parametric Oscillator (OPO) are directed towards the test cavity, and the reflected fields are detected with two homodyne measurements. A triangular mode cleaner cavity is used to spatially separate the entangled fields. An auxiliary Coherent Locking Field (CLF) phase modulated by an electro-optic modulator (EOM) is injected into the OPO to control the squeezing angle. \textbf{b,} Frequency spacing diagram with respect to the OPO resonances. The entangled signal and idler fields are generated within the OPO linewidth, separated by the OPO free spectral range. Fields within the pink (cyan) dotted box indicate fields that are detected by the signal (idler) homodyne. }
	\label{fig:Setup_supp}
\end{figure*}

	The EPR entangled source is a sub-threshold non-degenerate optical parametric oscillator (OPO) consisting of a periodically-poled potassium titanyl phosphate (PPKTP) nonlinear crystal embedded within a bow-tie cavity. The OPO is pumped with an optical field at frequency $2\omega_0 + \Delta$ generated from a second harmonic generator (SHG), and produces entangled signal and idler vacuum fields at frequency $\omega_0$ and $\omega_0+\Delta$ respectively \cite{Hage2010}, as shown in Figure \ref{fig:Setup_supp}. The OPO has a threshold pump power of 66.3\,mW and a signal-idler nominal separation of $\Delta = 851$\,MHz which corresponds to the OPO free spectral range.
	
	The signal and idler entangled states are injected into an over-coupled test cavity which consists of an input mirror with a 95\% power reflectivity, and a highly reflective end mirror. The test cavity has a length of L $\approx$ $0.5$\,m and a half-width-half-maximum (HWHM) of $\gamma_{tc}$ = 1.25\,MHz. The test cavity length is controlled using RF reflection locking \cite{Drever1983} with an orthogonally polarized locking field that is frequency-shifted from the main laser by $\Delta_{tc} \approx 152$\,MHz using a double passed Acousto Optic Modulator (AOM).  The relative detuning of the signal and idler field to the test cavity is controlled by fine-tuning $\Delta$ and $\Delta_{tc}$.
	
	The reflected fields of the test cavity are picked off using a Faraday isolator and are directed toward the measurement stage. The signal and idler fields are spatially separated with an impedance matched triangular mode cleaner cavity which allows the signal field to be transmitted and the idler field to be reflected. The mode cleaner has a free spectral range of $\sim 2\Delta$, and a HWHM of $\gamma_{MC}$ = 60\,MHz, well above the measurement band. The cavity has a measured transmission (reflection) of 97 (3)\.\%, limited predominately by mode matching loss.
	
	The signal and idler fields are detected at two separate balanced homodyne detectors which allow their quadrature to be measured at any readout angle $\phi$ by changing the phase of either the signal or idler local oscillator (LO). The signal and idler homodyne LOs are generated by phase modulating a tap-off of the main laser field with a waveguide EOM at modulation frequency $\Delta/2$. The modulation depth is chosen in order to maximise the power in the lower and upper 1st-order sidebands which correspond to the signal and idler LOs respectively. The field is split into two paths and is transmitted through two different optical cavity in order to isolate the signal and idler LOs. The optical cavities with a linewidth of the order of 800\,kHz also provided additional amplitude noise suppression, lowering the common-mode rejection requirement for the homodyne detectors. The homodyne readouts are combined passively with an electronic combiner where the final measurement spectrum is taken, and the power of the signal and idler LO is tuned to ensure that the readouts are combined equally.

\section{Output fields of a non-degenerate OPO} 
\label{OPO}

	The Hamiltonian for a non-degenerate OPO with perfect phase-matching is given by
\begin{align}
	H = \hbar (2\omega+\Delta)b^\dagger b + \hbar\omega a_s^\dagger a_s + \hbar(\omega+\Delta) a_i^\dagger a_i + \frac{i \hbar \epsilon}{2}\left(a_s^\dagger a_i^\dagger b-a_s a_i b^\dagger\right)
	\label{eq:Hamiltonian}
\end{align}
where $a_{s,i}, a_{s,i}^\dagger$ are the annihilation and creation operators of the intra-cavity signal and idler fields at frequencies $\omega$ and $\omega+\Delta$ respectively, $b, b^\dagger$ are the operators of the second harmonic intracavity pump field at frequency $2\omega+\Delta$, and $\epsilon$ is the non-linear coupling parameter. The output fields of the OPO can be calculated using the quantum Langevin equations \cite{Collett1984, Gardiner1985} . Assuming no pump field depletion, which is valid for the OPO operating below threshold, and that the cavity is held on resonance for the signal and idler fields, the equations of motion for the fundamental fields are
\begin{align}
	\dot{a}_s &= -\gamma_{tot} a_s + \frac{\epsilon}{2} |b| e^{i \theta_b} a_i^\dagger + \sqrt{2\gamma_{in}}A_{s,in} + \sqrt{2\gamma_{l}}A_{s,l} \\
	\dot{a}_i &= -\gamma_{tot} a_i + \frac{\epsilon}{2} |b| e^{i \theta_b} a_s^\dagger + \sqrt{2\gamma_{in}}A_{i,in} + \sqrt{2\gamma_{l}}A_{i,l} 
	\label{eq:EoM}
\end{align}
together with its complex conjugates. $A_{in}$ and $A_l$ are the vacuum fluctuations that couples into the OPO via the input coupler, and via intra-cavity losses respectively. $\gamma_{in}$ and $\gamma_l$ are the decay rates for the input coupler and the intra-cavity losses in the fundamental field, and $\gamma_{tot} = \gamma_{in} + \gamma_l$ is the total OPO cavity decay rate for the fundamental field. $\theta_b$ is the phase of the input pump field.

	Solving the 4 equations of motion in the frequency domain gives the following relation
\begin{align}
	\tilde{\mathbf{a}}\left(\Omega\right) &= \left(i\Omega \mathbf{I}-\gamma_{tot}\mathbf{M}\right)^{-1} \left[\sqrt{2\gamma_{in}}\tilde{\mathbf{A}}_{in}\left(\Omega\right) + \sqrt{2\gamma_{l}}\tilde{\mathbf{A}}_{l}\left(\Omega\right) \right]
	\label{eq:sol}
\end{align}
where
\begin{align}
	\tilde{\mathbf{a}} = 
	\begin{pmatrix}
	a_s \\
	a_s^\dagger \\
	a_i \\
	a_i^\dagger
	\end{pmatrix},
	\mathbf{M} = 
	\begin{pmatrix}
	-1 						& 0 							& 0 								& x e^{i \theta_b} \\
	0 							& -1 							&  x e^{-i \theta_b} 	& 0 \\
	0 							& x e^{i \theta_b} 	& -1 								& 0 \\
	x e^{-i \theta_b} & 0 							& 0 								& -1
	\end{pmatrix},
	\tilde{\mathbf{A}}_{in} = 
	\begin{pmatrix}
	A_{s,in} \\
	A_{s,in}^\dagger \\
	A_{i,in} \\
	A_{i,in}^\dagger
	\end{pmatrix},
	\tilde{\mathbf{A}}_{l} = 
	\begin{pmatrix}
	A_{s,l} \\
	A_{s,l}^\dagger \\
	A_{i,l} \\
	A_{i,l}^\dagger
	\end{pmatrix}
	\label{eq:sol2}
\end{align}
and $x=\frac{\epsilon |b|}{ 2 \gamma_{tot}}$ is the normalised pump parameter.

	The output of the OPO can be determined by using the following input-output relation
\begin{align}
	\tilde{\mathbf{A}}_{out}\left(\Omega\right) &= \sqrt{2\gamma_{in}}\tilde{\mathbf{a}}\left(\Omega\right) - \tilde{\mathbf{A}}_{in}\left(\Omega\right) \\
	&= \left(2\gamma_{in}\left(i\Omega\mathbf{I}-\gamma_{tot}\mathbf{M}\right)^{-1}-\mathbf{I}\right)\tilde{\mathbf{A}}_{in}\left(\Omega\right) + 2\sqrt{\gamma_l\gamma_{in}}\left(i\Omega\mathbf{I}-\gamma_{tot}\mathbf{M}\right)^{-1}\tilde{\mathbf{A}}_{l}\left(\Omega\right)
	\label{eq:out}
\end{align}

	The quadrature fluctuations $X_1$ (amplitude) and $X_2$ (phase) of these fields can be calculated using the transformation 
\begin{align}
	\tilde{\mathbf{X}}_{out} = \mathbf{\Gamma} \tilde{\mathbf{A}}_{out}
	\label{eq:trans}
\end{align}
where
\begin{align}
	\tilde{\mathbf{X}}_{out} = 
	\begin{pmatrix}
	X_{s,1,out} \\
	X_{s,2,out} \\
	X_{i,1,out} \\
	X_{i,2,out}
	\end{pmatrix},
	\mathbf{\Gamma} = 
	\begin{pmatrix}
	1 							& 1 							& 0 								& 0 \\
	-i 							& i 							&  0								& 0 \\
	0 							& 0						 	& 1 								& 1 \\
	0						    & 0 							& -i 								& i
	\end{pmatrix}
	\label{eq:trans2}
\end{align}
The quadrature fluctuations of the OPO output in terms of the input fields are then given by
\begin{align}
	\tilde{\mathbf{X}}_{out} = \mathbf{T}_{OPO} \tilde{\mathbf{X}}_{in} + \mathbf{T}_{OPO,l} \tilde{\mathbf{X}}_{l}
	\label{eq:output}
\end{align}
where
\begin{align}
	\mathbf{T}_{OPO}  &= \mathbf{\Gamma}\left(2\gamma_{in}\left(i\Omega\mathbf{I}-\gamma_{tot}\mathbf{M}\right)^{-1}-\mathbf{I}\right)\mathbf{\Gamma}^{-1} \\
	\mathbf{T}_{OPO,l}  &= \mathbf{\Gamma}2\sqrt{\gamma_l\gamma_{in}}\left(i\Omega\mathbf{I}-\gamma_{tot}\mathbf{M}\right)^{-1}\mathbf{\Gamma}^{-1}
	\label{eq:output2}
\end{align}

	The variances, $V_{(k)} = |X_{(k)}|^2$, of the output quadratures can be calculated by setting all the input fields to consist of vacuum fluctuations only, i.e. $V_{in} = 1$. This results in 
\begin{align}
	V_{out} = 1 + \frac{8x^2\eta_{esc}}{x^4 + 2x^2((\Omega/\gamma_{tot})^2-1) +((\Omega/\gamma_{tot})^2+1)^2}
	\label{eq:V}
\end{align}
for both the quadratures of the signal and idler fields, where $\eta_{esc} = \gamma_{in}/{\gamma_{tot}}$ is the OPO escape efficiency. The output variance does not go below the shot noise limit, i.e. $V_{out} > 1$ due to the sidebands about both the signal and idler fields being uncorrelated.

	However, by considering the combination $X_{cond} = X_{s,1,out} - X_{i,1,out}$, the variance of the combined state is
\begin{align}
	V_{cond} = V_{+} \cos^2\left(\theta_b/2\right) + V_{-} \sin^2\left(\theta_b/2\right)
	\label{eq:epr}
\end{align}
where
\begin{align}
	V_{\pm} = 2\left(1\pm \frac{4x\eta_{esc}}{(1\mp x)^2 + (\Omega/\gamma_{tot})^2}\right)
	\label{eq:epr2}
\end{align}
This results in a reduction below the shot noise limit for $\theta_b = \pi$, due to the correlations between the upper sidebands of the signal fields with the lower sidebands of the idler field, and vice-versa from the parametric down-conversion process. Equations \ref{eq:epr} and \ref{eq:epr2} are almost identical to the standard squeezed output equations from a degenerate OPO apart from the additional factor of two, which is due to the quantum noise level being referenced to the sum of the signal and idler shot noise level.

\section{Model for a detuned test cavity}

The equations of motion for a detuned test cavity is given by
\begin{align}
	\dot{\mathbf{a}}_{tc} &= \gamma_{tc}\mathbf{M_{tc}} \mathbf{a}_{tc} +  \sqrt{2\gamma_{tc,in}}\mathbf{A}_{tc,in} +  \sqrt{2\gamma_{tc,l}}\mathbf{A}_{tc,l} 
\end{align}
where
\begin{align}
	\mathbf{a}_{tc} = 
	\begin{pmatrix}
	a_{tc} \\
	a_{tc}^\dagger 
	\end{pmatrix},
	\mathbf{M_{tc}} = 
	\begin{pmatrix}
	-1 + i \Delta_{\gamma_{tc}}		& 0 				\\		
	0 															& -1 - i \Delta_{\gamma_{tc}}
	\end{pmatrix},
	\mathbf{A}_{tc,in/l} = 
	\begin{pmatrix}
	A_{tc,in/l} \\
	A_{tc,in/l}^\dagger 
	\end{pmatrix}
	\label{eq:tc_sol2}
\end{align}
$\mathbf{a}_{tc}$ is the creation and annihilation operators of the field inside the test cavity, $A_{tc,in}$ and $A_{tc,l}$ are the vacuum fluctuations that couples into the cavity via the input coupler, and via intra-cavity losses respectively. $\gamma_{tc,in}$ and $\gamma_{tc,l}$ are the decay rates for the input coupler and the intra-cavity losses, and $\gamma_{tc} = \gamma_{tc,in} + \gamma_{tc,l}$ is the total test cavity decay rate. $\Delta_{\gamma_{tc}}$ is the test cavity detuning normalized to $\gamma_{tc}$. 

Following the procedure in Section \ref{OPO}, the quadrature fluctuations on the test cavity reflection are
\begin{align}
	\tilde{\mathbf{X}}_{tc,out} = \mathbf{T}_{tc} \tilde{\mathbf{X}}_{tc,in} + \mathbf{T}_{tc,l} \tilde{\mathbf{X}}_{tc,l}
	\label{eq:TC_output}
\end{align}
where
\begin{align}
	\mathbf{T}_{tc}  &= \mathbf{\Gamma}\left(2\gamma_{tc,in}\left(i\Omega\mathbf{I}-\gamma_{tc}\mathbf{M_{tc}}\right)^{-1}-\mathbf{I}\right)\mathbf{\Gamma}^{-1} \\
	&=
	\begin{pmatrix}
	2\chi_c\eta_{tc}\left(1+i \Omega_{\gamma_{tc}}\right)-1	& -2\chi_c\eta_{tc}\Delta_{\gamma_{tc}}			\\		
	2\chi_c\eta_{tc}\Delta_{\gamma_{tc}}			& 2\chi_c\eta_{tc}\left(1+i \Omega_{\gamma_{tc}}\right)	-1
	\end{pmatrix} \\
	\mathbf{T}_{tc,l}  &= \mathbf{\Gamma}2\sqrt{\gamma_{tc,l}\gamma_{tc,in}}\left(i\Omega\mathbf{I}-\gamma_{tc}\mathbf{M_{tc}}\right)^{-1}\mathbf{\Gamma}^{-1} \\
	&= 2\chi_c\sqrt{\eta_{tc}(1-\eta_{tc})}
	\begin{pmatrix}
	1+i \Omega_{\gamma_{tc}}	& -\Delta_{\gamma_{tc}} \\
	\Delta_{\gamma_{tc}}								& 1+i \Omega_{\gamma_{tc}}
	\end{pmatrix}
	\label{eq:TC_output2}
\end{align}
with $\eta_{tc} = \gamma_{tc,in}/\gamma_{tc}$, $\Omega_{\gamma{tc}} = \Omega/\gamma_{tc}$ and $\chi_c^{-1} = \left(1+i \Omega_{\gamma_{tc}}\right)^2+\Delta_{\gamma_{tc}}^2$ is the optical susceptibility . The non-zero off-diagonal elements of $\mathbf{T}_{tc}$ when $\Delta_{\gamma_{tc}} \neq 0$ indicate that the input quadrature to the test cavity is allowed to couple into the orthogonal output quadrature, allowing the squeezed quadrature to be rotated.

\section{Derivation of the CLF error signals}

	The CLF error signal is generated by injecting a phase modulated field 
\begin{align}
	A_{CLF} \approx e^{i \phi_{c}} \left(A_{CLF,0} + A_{+} e^{\omega_m t} - A_{-} e^{-\omega_m t}\right)
	\label{eq:CLF}
\end{align}
into the OPO, where $\omega_m$ is the modulation frequency, $\phi_{c}$ is the phase of the CLF field,  $A_{CLF,0}$ is the carrier amplitude, and $A_{+,-}$ is the amplitude of the first order upper and lower phase modulated sidebands. The frequency of $A_{CLF,0}$ and $\omega_m$ is chosen such that only $A_{+,-}$ interacts with the OPO while $A_{CLF,0}$ is held on the OPOs anti-resonance.

	Similar to equation \ref{eq:sol}, the intra-cavity fields due to the CLF sideband injection is
\begin{align}
	\tilde{\mathbf{a}}\left(\Omega\right) &= \left(i\Omega \mathbf{I}-\gamma_{tot}\mathbf{M}\right)^{-1} \sqrt{2\gamma_{CLF}}\tilde{\mathbf{A}}_{CLF}
	\label{eq:sol_CLF}
\end{align}	
where
\begin{align}
	\tilde{\mathbf{A}}_{CLF} = 
	\begin{pmatrix}
	A_+ e^{i\phi_{c}} \\
	A_+ e^{-i\phi_{c}} \\
	-A_- e^{i\phi_{c}} \\
	-A_- e^{-i\phi_{c}}
	\end{pmatrix}
	\label{eq:sol_CLF2}
\end{align}
and $\gamma_{CLF}$ is the decay rate of the mirror where the CLF was injected. The reflected and transmitted fields are determined using the following input-output relations
\begin{align}
	\tilde{\mathbf{A}}_{CLF,r} &= \sqrt{2\gamma_{CLF}}\tilde{\mathbf{a}}- \tilde{\mathbf{A}}_{CLF} \\
	\tilde{\mathbf{A}}_{CLF,t} &= \sqrt{2\gamma_{in}}\tilde{\mathbf{a}}
	\label{eq:out_CLF}
\end{align}

The error signal on reflection is extracted by demodulating the reflected field at frequency $\omega_m$. By considering the static change to the error signal, i.e. setting $\Omega = 0$, the error signal is proportional to
\begin{align}
	E_{CLF,r} &= \operatorname{Im}\left(A_{+,r}A_{CLF,0}e^{-i\phi_c} + A_{-,r}^\dagger A_{CLF,0}e^{i\phi_c}\right) \\
	&\propto \frac{x}{x^2-1}\frac{\gamma_{CLF}}{\gamma_{tot}}\sin(\theta_b - 2\phi_c)
	\label{eq:CLF_refl}
\end{align}
which allows the CLF to be phase locked to the pump field.

On transmission, the upper CLF sideband propagates with the entangled fields and beats the idler LO at the homodyne detector. The error signal is extracted by demodulating the homodyne signal at the difference frequency between the fields ($\Delta_{CLF}$ in the main text), and is proportional to
\begin{align}
	E_{CLF,t} &= \operatorname{Im}\left(A_{+,t}A_{LO}e^{-i\phi_{LO}}\right) \\
	&\propto \frac{1}{x^2-1}\frac{\sqrt{\gamma_{CLF}\gamma_{in}}}{\gamma_{tot}} \left(x \sin(\theta_b - \phi_{LO} - \phi_c) + \sin(\phi_{LO}-\phi_c)\right)
	\label{eq:CLF_trans}
\end{align}
With the reflected CLF lock engaged, i.e. $\theta_b = 2\phi_c$, the transmitted error signal simplifies to
\begin{align}
	E_{CLF,t} \propto \sin (\theta_b/2 - \phi_{LO})
	\label{eq:CLF_trans2}
\end{align}
allowing the idler LO to be phase locked to the pump field. A similar error signal can also be derived between the lower CLF sideband and the signal LO.

\end{document}